\documentclass[pra,twocolumn]{revtex4}
\usepackage{amsmath,amssymb,graphicx}
\newcommand{\be}{\begin{equation}}
\newcommand{\ee}{\end{equation}}
\newcommand{\x}{{\bf x}}
\newcommand{\z}{{\bf z}}
\newcommand{\q}{{\bf q}}
\newcommand{\p}{{\bf p}}
\newcommand{\N}{\mathcal{N}}
\newcommand{\om}{{\pmb \omega}}

\begin{document}
\title{Semiclassical propagation of coherent states using complex and real trajectories}
\author{Marcel Novaes and Marcus A. M. de Aguiar}
\affiliation{Instituto de F\'{i}sica ``Gleb Wataghin",
Universidade Estadual de Campinas, 13083-970 Campinas, S\~ao
Paulo, Brazil}

\begin{abstract}
A semiclassical approximation to the time evolution of coherent
states may be derived from a saddle point approximation to the
exact quantum propagator, and in general it involves complex
classical dynamics. We generalize previous one-dimensional results
to $d$ dimensions, and for the case $d=2$ we present several
applications. We also consider other simple approximations that
depend only on real classical trajectories, but are not initial
value representations. These approximations are able to reproduce
interference and tunnelling effects, and involve propagating a few
classical initial conditions compatible with the quantum
uncertainties.

\end{abstract}

\maketitle

\section{Introduction}
Semiclassical propagators involving complex classical trajectories
in real time have appeared in the coherent states representation
around 25 years ago \cite{klauder,weiss}. A stationary phase
approximation to the transition amplitude $\langle
z'|e^{-iHT/\hbar}|z\rangle$, where $|z\rangle$ is a coherent
state, leads to trajectories satisfying the usual Hamilton
equations subject to special boundary conditions that can only be
satisfied in a complexified phase space. Numerical calculations in
this representation have been done both for chaotic systems
\cite{adachi,cabelo} and for one dimensional systems
\cite{marcus,xavier,parisio} (see also \cite{voorhis}; reviews can
be found in \cite{rubin,baranger}). Semiclassical calculations
involving complex trajectories in the mixed representation
$\langle x|e^{-iHT/\hbar}|z\rangle$, on the other hand, were
introduced in \cite{huber} and recently rediscovered \cite{aguiar}
for the one-dimensional case (see \cite{boiron} for a different
approach). Since the mixed representation is the most interesting
for the propagation of wave packets, our purpose here is to
generalize this formalism to many dimensions and to present some
applications.

The calculation of complex trajectories involves two difficulties:
first, the effective dimensionality of the phase space is doubled,
since both real and imaginary parts of position and momenta have
to be computed; second, the trajectories must satisfy mixed
boundary conditions, part at the initial time and part at the
final time, a problem known as `root search'. Therefore we also
consider the possibility of employing only real trajectories in
the semiclassical approximation. This is done by approximating the
complex trajectories by real ones, satisfying modified boundary
conditions that are less restrictive than the original ones.
Although such real trajectories approximations are always less
accurate than the original complex one, they are much simpler and
sometimes have practically the same accuracy \cite{aguiar}. Our
approach, both with complex and real trajectories, does not
involve integrations over initial conditions, a procedure that is
common to IVR -- Initial Value Representations (recent reviews of
this method can be found in \cite{review}). IVR methods are
usually easy to apply and reasonably accurate for long times.
Nevertheless, for short times the present method provides a much
clearer physical picture since only a few families of trajectories
are required.

We start from a coherent state $|\z\rangle$, where \be \z=
\frac{1}{\sqrt{2}}\left(B^{-1}\q+iC^{-1}\p\right),\ee and the
$d$-dimensional vectors $\q$ and $\p$ are the average values of
position and momentum for this state. The diagonal matrices $B$
and $C$ contain the position and momentum uncertainties,
respectively, and satisfy the condition $B=\hbar C^{-1}$. The
position representation of this coherent state is a Gaussian, \be
\langle
\x|\z\rangle=\mathcal{N}\exp\left\{\frac{i}{\hbar}\p^T(\x-\q)-\frac{1}{2}(\x-\q)^TB^{-2}(\x-\q)
\right\},\ee where
$\mathcal{N}=|B|^{-\frac{1}{2}}\pi^{-\frac{d}{4}}$ (we use the
symbol $|\cdot|$ for the determinant). After a time $T$, the
propagated wave function is given by \be
\label{wave}\psi(\x,T)=\langle \x|K(T)|\z\rangle, \ee where
$K(T)=e^{-iHT/\hbar}$.

In order to calculate the wave function semiclassically we shall
follow the procedure of \cite{huber,aguiar}. We first insert in
(\ref{wave}) a resolution of unity to obtain
\be\label{integ}\psi(\x,T)=\int d\x'\langle
\x|K(T)|\x'\rangle\langle \x'|\z\rangle, \ee and substitute the
quantity $\langle \x|K(T)|\x'\rangle$ by its semiclassical
Van-Vleck expression \cite{vleck,gutzwiller}. Then we make the
integration by the stationary exponent approximation. We shall see
that the stationary points are in general complex numbers, and
thus a deformation of the integration contour into the complex
plane is unavoidable, taking the classical trajectories involved
in the approximation to a complex phase space.

The Van-Vleck formula in $d$ dimensions is \be \label{VV}\langle
\x|K(T)|\x'\rangle_{VV}=(2\pi
i\hbar)^{-d/2}\sqrt{\left|-S_{\x\x'}\right|}e^{\frac{i}{\hbar}S},\ee
where $S(\x,\x',T)$ is the action of the classical trajectory that
goes from $\x'$ to $\x$ in time $T$ and $S_{\x\x'}$ is the matrix
of its second derivatives (we have incorporated Morse phases in
$S$). If more than one such trajectory exists, one should sum
their contributions. Before performing the integration, let us
express the determinant in (\ref{VV}) in terms of the elements of
the tangent matrix. As shown in the appendix, \be\label{pref}
\langle \x|K(T)|\x'\rangle_{VV}=\frac{(2\pi
i)^{-d/2}}{|B|\sqrt{|M_{\x\p}|}}e^{\frac{i}{\hbar}S}.\ee

Note that in the position representation nothing is said about the
momentum of the corresponding classical trajectory, and therefore
it is not necessary to introduce any complexification. In the
coherent states representation the boundary conditions are too
stringent as one tries to specify not only the initial and final
points (and the time) but also the initial and final momenta
\cite{baranger}.

In the next section we shall calculate the integral (\ref{integ})
in the saddle point approximation, valid in the semiclassical
limit. In section III we develop further approximations that
involve only real classical trajectories. We show some
illustrative numerical applications in section IV and present our
conclusions in section V.

\section{Complex trajectories}
In the semiclassical limit the wave function in (\ref{integ}) can
be written as \be\label{psi} \psi(\x,T)=\frac{\N}{|B|}(2\pi
i)^{-d/2}\int
d\x'\frac{\exp\{\frac{i}{\hbar}\Phi(\x,\x',T)\}}{\sqrt{|M_{\x\p}|}},\ee
where \be \frac{i}{\hbar}\Phi=\frac{i}{\hbar}[S+{\bf p}^T({\bf
x'}-{\bf q})]-\frac{1}{2}({\bf x'}-{\bf q})^TB^{-2}({\bf x'}-{\bf
q}).\ee We evaluate this integral by the usual saddle point
method, which consists in expanding the exponent to second order
around its stationary point ${\bf x}'_0$, while the prefactor is
simply evaluated at this point. After performing the resulting
Gaussian integration, this leads to the semiclassical
approximation \be \label{integ2}\psi_{\rm
sc}(\x,T)=\frac{\N}{|B|}(2\pi i)^{-d/2}
\frac{\exp\{\frac{i}{\hbar}\Phi_0\}}{\sqrt{|M_{\x\p}|}}\sqrt{-\frac{(2\pi)^d}{|\Phi_{\x'\x'}|}},\ee
where \be \Phi_{\x'\x'}=\frac{i}{\hbar}S_{\x'\x'}-B^{-2}\ee and
$\Phi_0=\Phi(\x,\x'_0,T)$.

The stationary point $\x'_0$, determined imposing the condition
$\left.\nabla'\Phi\right|_{\x'_0}=0$, is given by \be\label{stat}
B^{-1}(\x'_0-\q)=iC^{-1}(\p-\p'_0),\ee where
$\p'_0(\x,\x'_0,T)=-\left.\nabla' S\right|_{\x'_0}$. Both $\p'_0$
and $\x'_0$ are in general complex numbers, and the whole
classical trajectory therefore takes place in a complex phase
space. It leaves $\x'_0$ at time $0$ with the complex momentum
$\p'_0$ and arrives at the real position $\x$ at time $T$. The
classical action $S$ and the tangent matrix $M$ will also be
complex in general.

Using the relation (see appendix) \be
\label{det}|\Phi_{\x'\x'}|=\frac{i^d}{|B|^2}\frac{|M_{\x\x}+iM_{\x\p}|}{|M_{\x\p}|},\ee
the final result may be written in terms of the tangent matrix as
\be\label{final}\psi_{\rm
sc}(\x,T)=\frac{\N(-i)^{d-1}}{\sqrt{|M_{\x\x}+iM_{\x\p}|}}
\exp\{\frac{i}{\hbar}\Phi_0\}. \ee This generalizes the one
dimensional formula presented in \cite{huber,aguiar}, to which it
reduces for separable systems and that has proven to be accurate
in the evolution of wave packets in many different systems. It is
of course exact for the propagation of a $d$-dimensional coherent
state in free space and in potentials up to quadratic (harmonic
oscillator, charged particle in constant
electromagnetic/grativational field). Differently from the Van
Vleck approximation, the prefactor involves the square root of a
complex number (remember that $|\cdot|$ is a determinant, not a
modulus), and its phase must be determined dynamically with the
condition that for $T=0$ we have $M_{\x\x}=1$ and $M_{\x\p}=0$.

The semiclassical approximation (\ref{final}) depends on complex
trajectories $(\q(t),\p(t))$ satisfying the boundary conditions
\be \frac{1}{\sqrt{2}}\left(B^{-1}\q(0)+iC^{-1}\p(0)\right)={\bf
z},\quad \quad \q(T)=\x,\ee where we have used the fact that
(\ref{stat}) can be written
$B^{-1}\x'_0+iC^{-1}\p'_0=B^{-1}\q+iC^{-1}\p$. The final value of
the momentum is not restricted and will be complex in general.
Following Klauder and Adachi \cite{klauder,rubin,adachi} we may
write the initial condition as \be \q(0)=\q+\om, \quad
\p(0)=\p+iCB^{-1} \om, \ee where $\om={\pmb\alpha}+i{\pmb\beta} $
is a complex vector to be determined. The first condition is
automatically satisfied for any $\om$. For a fixed time $T$ the
propagation of this complex initial condition defines a complex
map $\om\to\q(T)$, the properties of which have been studied in
detail for the one-dimensional case in \cite{rubin}. Only for some
values of $\om$ will it happen that $\q(T)\in \mathbb{R}^d$, and
we denote the set of all those points by $\Omega$. It is easy to
see that $\om=0$ belongs to $\Omega$, in which case we have the
classical trajectory of the center of the wave packet.

However, the inverse of the map $\om\to\q(T)$ is in general
globally multivalued: there may be many trajectories that end at
the same $\q(T)$. Therefore $\Omega$ will consist in a finite
collection of $d$-dimensional disjoint sets, called families. In
the vicinity of a critical point (i.e., one for which $\partial
\q(T)/\partial\om$ is zero) the map is two-to-one, provided the
second derivative is not zero. Such a critical point is also
called a phase space caustic. At these points
$|M_{\x\x}+iM_{\x\p}|\to 0$, thus preventing the validity of the
semiclassical calculation. It is possible to develop a
semiclassical approximation based on the Airy function that
remains valid near caustics. For the one-dimensional case this has
been derived in \cite{uniform}.

The family of trajectories that contains the point $\om=0$ is
called the main family, and it provides the most important
contribution to the semiclassical approximation. As time
increases, other families may become relevant. The imaginary part
of $\Phi_0$ is positive for all trajectories that belong to the
main family, but this may not be the case for other families. When
$\rm{Im}(\Phi_0)<0$ one has a contribution that diverges when
$\hbar\to 0$ and therefore must be discarded. These are called
non-contributing trajectories, and for some families it is
necessary to introduce a cut-off in order to avoid them.

Another delicate point is that of Stokes lines and exponential
dominance, which is intrinsic to many asymptotic formulations. In
the usual one-dimensional WKB, for example, the semiclassical
approximation for stationary states becomes singular at classical
turning points, and one must connect different local solutions by
an analytic continuation. In so doing, one finds that there must
be a change in the number of contributions along certain lines
called Stokes lines \cite{stokes,dingle,berry,shudo}. In the
vicinity of such a line one contribution dominates exponentially
over the other, and one is free to place a cut-off (the error due
to the cut-off is less than the error due to the semiclassical
approximation). The same phenomenon appears in the present
formalism. Even though the location of these lines is hard to
determine in principle, in practice when crossing them there
appears a false divergence in the approximation, which can be
easily detected \cite{aguiar}.

\section{Approximations based on real trajectories}
One may wish to find approximations for the expression
(\ref{final}) that involve only real trajectories. There are many
such possibilities. One possible choice is the trajectory that
starts at the real point $\q$ with initial momentum $\p_i$,
different from $\p$, and after a time $T$ arrives at $\x$. Another
possibility is a trajectory that starts with momentum $\p$ but
from a different point $\q_i$ and also arrives at $\x$. We can
also give up the final point condition, for example, by choosing
the unique trajectory that starts at $\q$ with momentum $\p$. All
these possibilities are similar to the ones already existent in
the one-dimensional case \cite{aguiar}, but in more dimensions one
can in principle come up with others. For example, in two
dimensions a trajectory may exist that starts at $(q_x,q_{yi})$
with momentum $(p_{xi},p_y)$ and ends at $(x,y)$, but with
$q_{yi}\neq q_y$ and $p_{xi}\neq p_x$. All these real trajectories
should be good approximations for the complex stationary
trajectory if the latter is not too deep into the complex plane.

An important method that is also based on real trajectories is the
`cellular dynamics', developed by Heller \cite{cellular,houches},
in which a grid of initial conditions is evolved and each
contribution to the propagator is obtained by a linearization of
the dynamics. This method was initially used to propagate wave
packets \cite{cellular} and later to obtain coherent state
correlation functions $\langle z'|e^{-iHT/\hbar}|z\rangle$ in
chaotic systems \cite{celchaos1,celchaos2}. The calculations we
present in this section are close in spirit to these works, but
instead of following the `cellular' approach we start from the
complex trajectory approximation (\ref{final}), and we also
consider a variety of boundary conditions that the real
trajectories may satisfy. Using different boundary conditions we
obtain the `central' trajectory approximation \cite{houches} and
also more general results similar to the `off-centered' one
presented in \cite{celchaos2}.

This section regards only calculation of wave functions, but a
discussion of the quantity $\langle z'|e^{-iHT/\hbar}|z\rangle$
that proceeds along the same lines may be found in \cite{realzz}.

\subsection{Approximation via central trajectory}
The classical trajectory that starts at $(\q,\p)$ will end, after
a time $T$, in the point $(\q_r,\p_r)$. Following \cite{aguiar} we
write
\begin{eqnarray}
\x'_0&=&\q+\Delta\x'\\\label{dp}\p'_0&=&\p+\Delta\p'=\p-S_{\x'\x'}\Delta\x'-S_{\x'\x}\Delta\x\\
\x&=&\q_r+\Delta\x\\\p&=&\p_r+\Delta\p=\p_r+S_{\x\x'}\Delta\x'+S_{\x\x}\Delta\x.
\end{eqnarray} The stationary exponent condition can be written as
$\Delta\p'=i\hbar B^{-2}\Delta\x'$, and equation (\ref{dp}) can be
solved to give (see appendix) \be\label{x2x}
\Delta\x'=B(M_{\x\x}+iM_{\x\p})^{-1}B^{-1}\Delta\x.\ee

Now we expand the exponent in (\ref{final}) around this trajectory
to second order in $\Delta\x$. The expansion of the action is
\begin{align} S\simeq S_r+&\p_r^T\Delta\x-\p^T\Delta\x'\nonumber\\&+\frac{1}{2}(\begin{array}{cc}
  \Delta \x & \Delta \x'
\end{array})\begin{pmatrix}
  S_{\x\x} & S_{\x\x'} \\
  S_{\x'\x} & S_{\x'\x'}
\end{pmatrix}\left(\begin{array}{c}
  \Delta \x \\
  \Delta \x'
\end{array}\right). \end{align} The remaining terms are simply
$\p^T\Delta\x'$, which cancels out, and
$\Delta\x'^TB^{-2}\Delta\x'$. In the quadratic terms we introduce
the tangent matrix and use (\ref{x2x}) to obtain
\be\label{heller}\psi_{\q\p}(\x,T)=\frac{\N(-i)^{d-1}}{\sqrt{|M_{\x\x}+iM_{\x\p}|}}
\exp\{\frac{i}{\hbar}\Phi_r\}, \ee where the exponent is given by
(see appendix) \be\label{helphi}
\frac{i}{\hbar}\Phi_r=\frac{i}{\hbar}(S_r+\p_r^T\Delta\x)-\frac{1}{2}\Delta\x
B^{-1}\Xi B^{-1}\Delta\x,\ee where
$\Xi=(M_{\p\p}-iM_{\p\x})(M_{\x\x}+iM_{\x\p})^{-1}$. Note that
this is always Gaussian in $\Delta\x$, with variable width.
Therefore this approximation can never account for interferences
or tunnelling effects. Notice that while the formula (\ref{final})
involves an infinite number of classical trajectories, at least
one for each value of $\x$, the one we just derived requires only
the trajectory that starts in $(\q,\p)$. For this reason this is
called an Initial Value Representation (IVR).

This formula was first derived by Heller \cite{heller} (see also
\cite{houches}) and is called the Thawed Gaussian Approximation or
TGA (it was rederived with some detail in \cite{baranger}). It
becomes exact in the semiclassical limit $\hbar\to 0$ (for a fixed
value of time) and has been used, for example, in the study of
decoherence \cite{fiete} and of scars in quantum chaotic systems
\cite{scars}. In the applications presented here we are interested
in quantum effects that cannot be reproduced by the TGA, and thus
we do not consider it any further.

\subsection{Approximation via trajectory ${\bf q}\to{\bf x}$}
Let us fix the initial coordinate of the trajectory, $\q$, and
demand that after a time $T$ it arrives at $\x$. We need to find
the initial momentum $\p_i$ for such trajectories, and in fact
there may be more than one that satisfy the above conditions. We
write
\begin{eqnarray}
\x'_0&=&\q+\Delta\x'\\\label{pes}\p'_0&=&\p_i+\Delta\p'=\p_i-S_{\x'\x'}\Delta\x'.
\end{eqnarray} Note that the complete expansion of $\p'_0$ to
first order should be
$\p_i-S_{\x'\x'}\Delta\x'-S_{\x'\x}\Delta\x$, but we are keeping
$\x$ fixed.

Equation (\ref{stat}) gives \be
B^{-1}\Delta\x'=iC^{-1}(\Delta\p-\Delta\p'),\ee where
$\Delta\p=\p-\p_i$. Using (\ref{pes}) we find \be
(\frac{i}{\hbar}S_{\x'\x'}-B^{-2})\Delta\x'=\Phi_{\x'\x'}\Delta\x'=-\frac{i}{\hbar}\Delta\p,\ee
which we can invert to write \be
\Delta\x'=-\frac{i}{\hbar}\Phi_{\x'\x'}^{-1}\Delta\p.\ee We now
expand the exponent in (\ref{final}) around this trajectory to
second order in $\Delta\x'$. Proceeding analogously to the
one-dimensional case \cite{aguiar} we obtain \be\label{qx0}
\frac{i}{\hbar}\Phi_{\q}=\frac{i}{\hbar}S_{\q}+\frac{1}{2\hbar^2}\Delta\p^T\Phi_{\x'\x'}\Delta\p,\ee
which, with a few algebraic manipulations, may be expressed in
terms of the tangent matrix (see appendix) as \be
\frac{i}{\hbar}\Phi_{\q}=\frac{i}{\hbar}S_{\q}-\frac{i}{2}\Delta\p^T
C^{-1}(M_{\x\x}+iM_{\x\p})^{-1}M_{\x\p}C^{-1}\Delta\p.\ee The wave
function becomes \be\label{qtox}
\psi_\q(\x,T)=\frac{\N(-i)^{d-1}}{\sqrt{|M_{\x\x}+iM_{\x\p}|}}
\exp\{\frac{i}{\hbar}\Phi_\q\}.\ee

The exponent contains the real action $S_\q$ and a term which is
Gaussian in the difference between $\p_i$, the initial momentum of
the trajectory, and $\p$, the average momentum of the initial
coherent state. It is important to note that $\p_i$ usually
depends on $\x$ in a complicated manner, and thus the final wave
packet will not, in general, be Gaussian. Also, there may exist
more than one value of $\p_i$, and a sum over all possible
trajectories would be required, resulting in interference terms.
Since the trajectory involved in the calculation depends on the
initial $\q$ and final $\x$ points, this is not an IVR.

\subsection{Approximation via trajectory ${\bf p}\to{\bf x}$}
We now fix the initial momentum of the trajectory and allow it to
start from a point $\q_i$ that is different from the center of the
wave packet. We write
\begin{eqnarray}
\x'_0&=&\q_i+\Delta\x'\\\p'_0&=&\p+\Delta\p'=\p-S_{\x'\x'}\Delta\x',
\end{eqnarray}
and use the stationary exponent condition (\ref{stat}) to find \be
(\frac{i}{\hbar}S_{\x'\x'}-B^{-2})\Delta\x'=-B^{-2}\Delta\q \ee or
$\Delta\x'=-\Phi_{\x'\x'}^{-1}B^{-2}\Delta\q$, with
$\Delta\q=\q-\q_i$. Once again, we expand the exponent in
(\ref{final}) to second order in $\Delta\x'$, but this time we
write it in terms of $\Delta\q$. The final result is
\be\label{ptox}
\psi_\p(\x,T)=\frac{\N(-i)^{d-1}}{\sqrt{|M_{\x\x}+iM_{\x\p}|}}
\exp\{\frac{i}{\hbar}\Phi_\p\},\ee where
\begin{align}
\frac{i}{\hbar}\Phi_\p&=\frac{i}{\hbar}(S_\p-\p^T\Delta\q)\nonumber\\
&-\frac{1}{2}\Delta\q^T
B^{-1}(M_{\x\x}+iM_{\x\p})^{-1}M_{\x\x}B^{-1}\Delta\q.\end{align}

We have obtained a Gaussian again, this time in the difference
between $\q_i$, the initial position of the trajectory, and $\q$,
the average position of the initial coherent state. This is again
not an IVR, and after a time $T$ it will not result in a Gaussian
in $\x$. It may as well display interference between different
existent classical trajectories.

\subsection{Approximation via a mixed trajectory}
We now restrict ourselves to a two-dimensional system and choose
the real trajectory that starts at $(q_x,q_{yi})$ with momentum
$(p_{xi},p_y)$ and ends at $(x,y)$, but with $q_{yi}\neq q_y$ and
$p_{xi}\neq p_x$. This time we have mixed conditions, and we set
\begin{eqnarray} x'_0&=&q_x+\Delta x',\\y'_0&=&q_{yi}+\Delta
y',\\\label{pxmix}p'_{x0}&=&p_{xi}+\Delta
p'_x=p_{xi}-S_{x'x'}\Delta x'-S_{x'y'}\Delta
y',\\\label{pymix}p'_{y0}&=&p_{y}+\Delta p'_y=p_y-S_{y'x'}\Delta
x'-S_{y'y'}\Delta y'.
\end{eqnarray} Using these equations, the stationarity conditions can
be cast in the form $\Delta\x'=-\Phi_{\x'\x'}^{-1}\Delta{\pmb
\xi}$, where \be \Delta{\pmb \xi}=\begin{pmatrix}
  i\Delta p_x/\hbar \\
  \Delta y/b_y^2
\end{pmatrix}, \quad \Delta\x'=\begin{pmatrix}
 \Delta x' \\
  \Delta y'
\end{pmatrix}.\ee

The expansion of the exponent to second order in $\Delta\x'$ is
$S\approx
S_M+\nabla'S^T\Delta\x'+\frac{1}{2}\Delta\x'^TS_{x'x'}\Delta\x'$
for the action, $\p^T(\x'_0-\q)=-\nabla'S^T\Delta\x'-p_y\Delta y$
for the term involving the wave packet momentum and $({\bf
x}'_0-{\bf q})^TB^{-2}({\bf x}'_0-{\bf
q})=\Delta\x'^TB^{-2}\Delta\x'+b_y^{-2}\Delta y(\Delta y-2\Delta
y')$ for the quadratic term. The linear terms in $\Delta\x'$
cancel, and after we change from $\Delta\x'$ to $\Delta\xi$ we
have \be \frac{i}{\hbar}\Phi_M=\frac{i}{\hbar}(S_M-p_y\Delta y
)-\frac{1}{2}\Delta\xi^T\Phi^{-1}_{\x'\x'}\Delta\xi-\frac{\Delta
y^2}{2b_y^2}, \ee where $S_M$ is the action for the mixed
condition trajectories. Once again this can be written in terms of
the tangent matrix, as we show in the appendix. The wave function
becomes \be \label{mist}\psi_M(x,y,T)=\frac{-i}{\sqrt{\pi
b_xb_y}}\frac{\exp\{\frac{i}{\hbar}\Phi_M\}}{\sqrt{|M_{\x\x}+iM_{\x\p}|}}.\ee

\subsection{Alternative derivation}
Given the integral in (\ref{psi}) one may argue that, if the
position uncertainties are very small, only the region around $\q$
will be relevant for the integration. Expanding the action to
second order around this point we have \begin{align}
\frac{i}{\hbar}\Phi\approx&
\frac{i}{\hbar}S(\x,\q)-\frac{i}{\hbar}(\p_i-\p)^T(\x'-\q)\nonumber\\&
+\frac{1}{2}(\x'-\q)^T\Phi_{\x'\x'}(\x'-\q),\end{align} where
$\p_i=-\left.\nabla'S\right|_{\x'=\q}$ is the initial momentum,
generally different from $\p$. Proceeding with the integration, we
find the same result as in section V.A, which is based on the
trajectory that starts at $\q$ with momentum $\p_i$ and ends at
$\x$ with any momentum at time $T$.

Jalabert and Pastawski \cite{jalabert} have used a similar
argument in their treatment of the quantum fidelity \be\label{fid}
\int \psi^{\ast}(\x,T)\psi_V(\x,T)d\x,\ee (in this equation
$\psi(\x,T)$ and $\psi_V(\x,T)$ are obtained from an initial wave
function by evolving it with two different Hamiltonians), but they
expanded the action to first order in the difference $\x'-\q$ (the
same procedure was used in \cite{petitjean}). After changing the
integration variable in (\ref{fid}) from $\x$ to $\p_i$, Vanicek
and Heller \cite{vanicek} arrive at a semiclassical result that is
free of caustics. Even though expanding the action to first order
only is inaccurate for simple systems such as the free particle
and the harmonic oscillator, their final formula seems to work
well in practice. The expansion to second order we just presented
is in principle more accurate, but the result it gives for the
semiclassical fidelity is sensitive to caustics and thus probably
less stable in numerical calculations.

Finally we note that, if instead of inserting a position
representation of unity in (\ref{wave}), we used a momentum
representation, \be \psi(\x,T)=\int d\p'\langle
\x|K(T)|\p'\rangle\langle \p'|\z\rangle,\ee then after a similar
second order expansion of the action, this time around $\p'=\p$,
we would arrive at the expression (\ref{ptox}) for the wave
function. This is justified when the momentum uncertainties are
small. The TGA approximation can also be obtained this way: one
must enforce a stationary phase condition on the imaginary part of
$\Phi(\x,\x')$ alone.

\section{Applications}
In this section we present a few numerical applications of the
approximations we have just derived. We compare the
complex-trajectories formula (\ref{final}) with the ones based on
real trajectories, and also with exact quantum mechanical
calculations, which we have carried out using Fast Fourier
Transform methods. The purpose here is not to obtain extremely
accurate numerical results, even though sometimes this is the
case, but rather to illustrate the usefulness of semiclassical
calculations in many different situations.

\begin{figure}[b]
\includegraphics[scale=0.29,angle=-90]{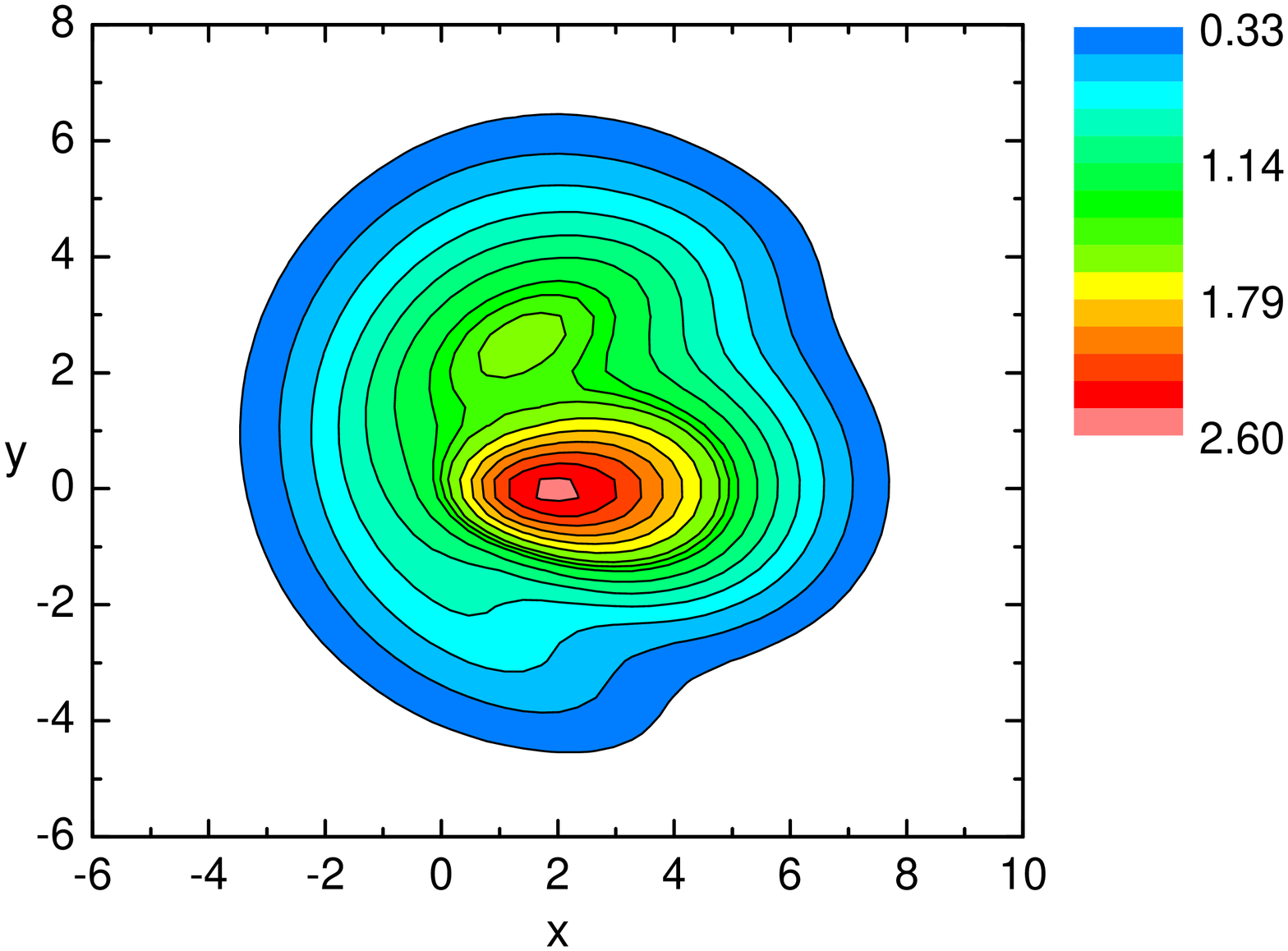}\\
\includegraphics[scale=0.29,angle=-90]{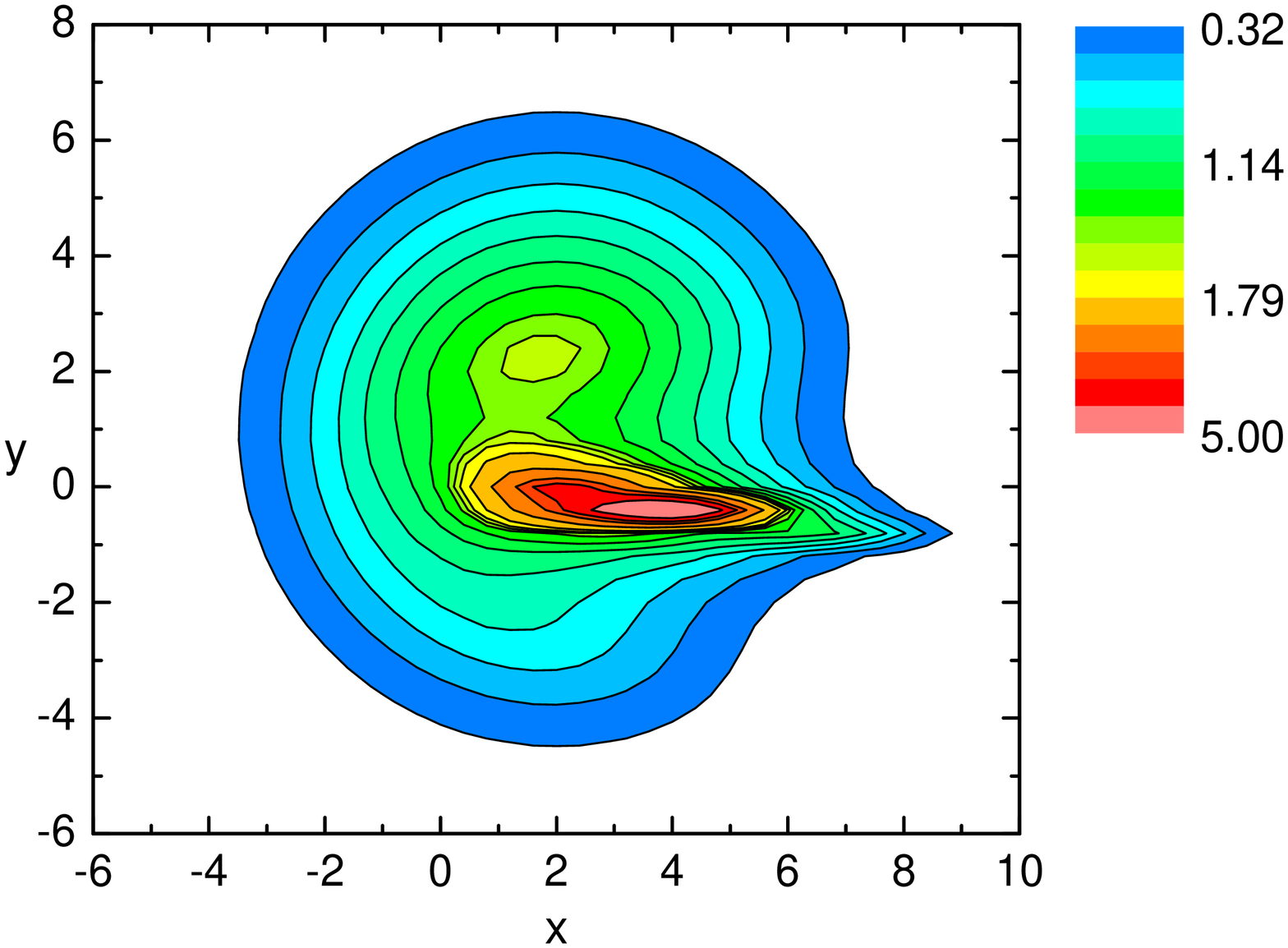}
\caption{(color online) Exact (up) and semiclassical (down)
probability densities (times $10^2$) at $T=4$, with $\q=(-10,1)$
and $\p=(3,0)$, in the case of an attractive Gaussian potential.
Except for the main peak (notice the change in scale), the wave
function is accurately reproduced, and $|\langle \psi|\psi_{\rm
sc}\rangle|^2\approx 92\%$}
\end{figure}

\subsection{Attractive Gaussian potential}

We start by investigating the semiclassical propagation in a two
dimensional attractive Gaussian potential, \be V(r)=-e^{-r^2},\ee
where $r^2=x^2+y^2$. A one dimensional version of this problem was
already considered in \cite{aguiar}, where the semiclassical
approximation was shown to be very accurate. This potential is
also interesting because, unless the particle's momentum is very
low (which is not the case we are interested in), there is only
one classical trajectory that contributes to the real
semiclassical formulas presented in sections III-A to III-D. In
the complex case there may be more than one trajectory, but we
will confine ourselves to the main family only, since it already
gives a very good result.

We place the wave packet initially at $\q=(-10,1)$, and chose
$b_x=b_y=1$, so that the impact parameter is equal to the wave
packet width. After a time interval of $t=4$ the main peak has
followed a curved trajectory, arriving at a negative value of $y$,
and a smaller peak appears around $y\approx 2$, as we can see in
Fig 1a. This is accurately reproduced by the semiclassical
approximation $\psi_{\rm sc}(\x,t)$, shown in Fig. 1b. The
secondary peak is recovered almost exactly, but the height of the
main peak is wrong by a factor of $2$ (notice the particular scale
that has been used). The phase of the wave function is also
recovered, and in fact the overlap \be\label{over} |\langle
\psi|\psi_{\rm sc}\rangle|^2 \ee is around $92\%$. It is important
to notice that the discrepancy comes from a small region around
the peak, and that the functions agree very well at all other
points. We have also calculated the real trajectory approximations
$\psi_{\q}(\x,t)$, $\psi_{\p}(\x,t)$ and $\psi_{M}(\x,t)$, but
none of them can be distinguished from the complex one at this
scale.

The erroneous increase in the main peak is probably due to the
presence of a caustic in complex phase space. Even though only one
real trajectory exists, in the complex case there may be more than
one, leading to critical points in the map described in section
II. In order to obtain a better approximation in the vicinity of
the peak, either a uniform approximation or incorporation of this
secondary family of trajectories would be necessary. Finding these
trajectories in practice is the notorious root-search problem,
known to be very difficult in more than one dimension. The
accuracy of the simpler formulas (\ref{qtox}), (\ref{ptox}) and
(\ref{mist}) in this case shows that they can be of practical use.

\begin{figure}[t]
\includegraphics[scale=0.29,angle=-90]{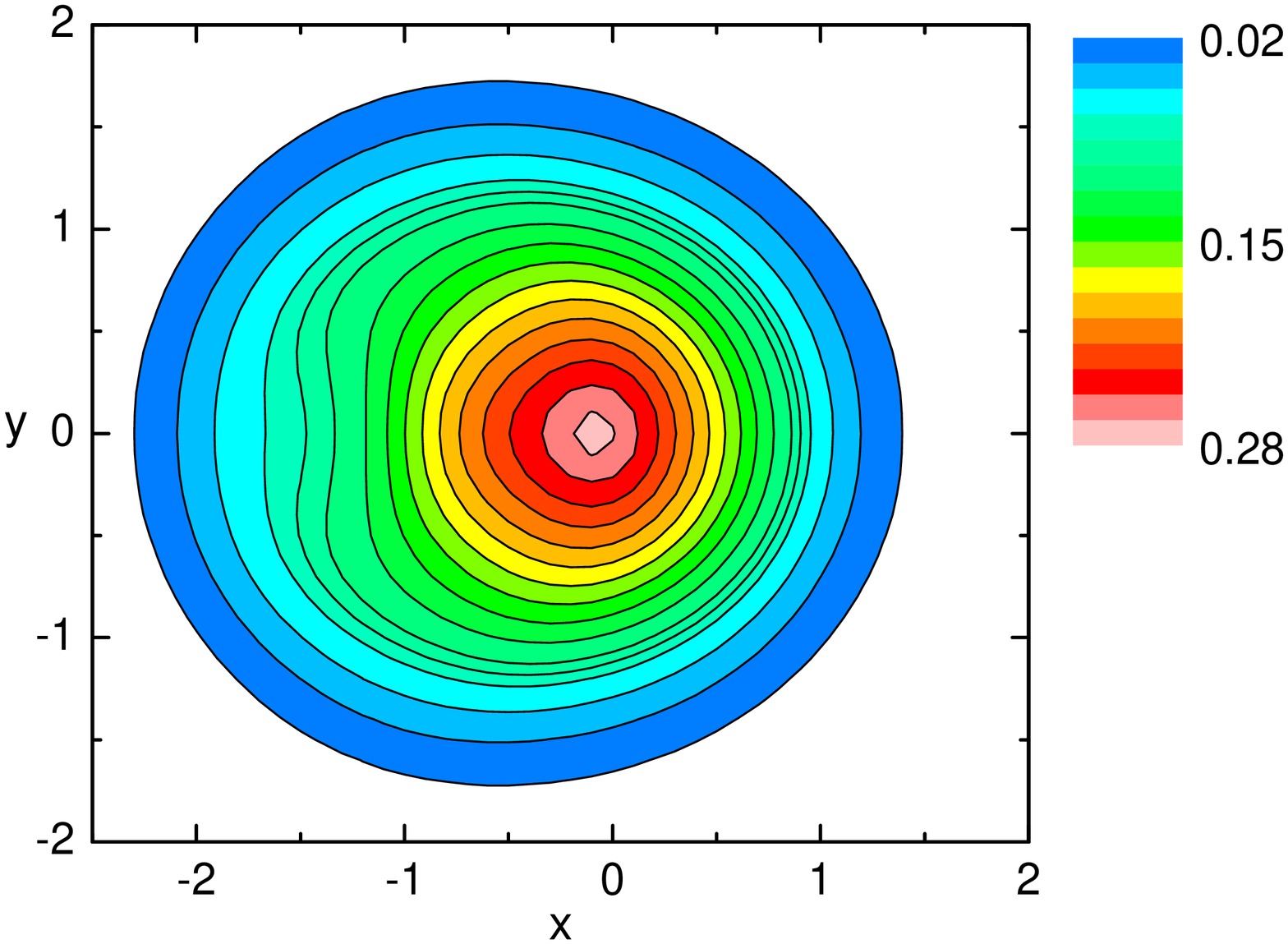}\\
\includegraphics[scale=0.29,angle=-90]{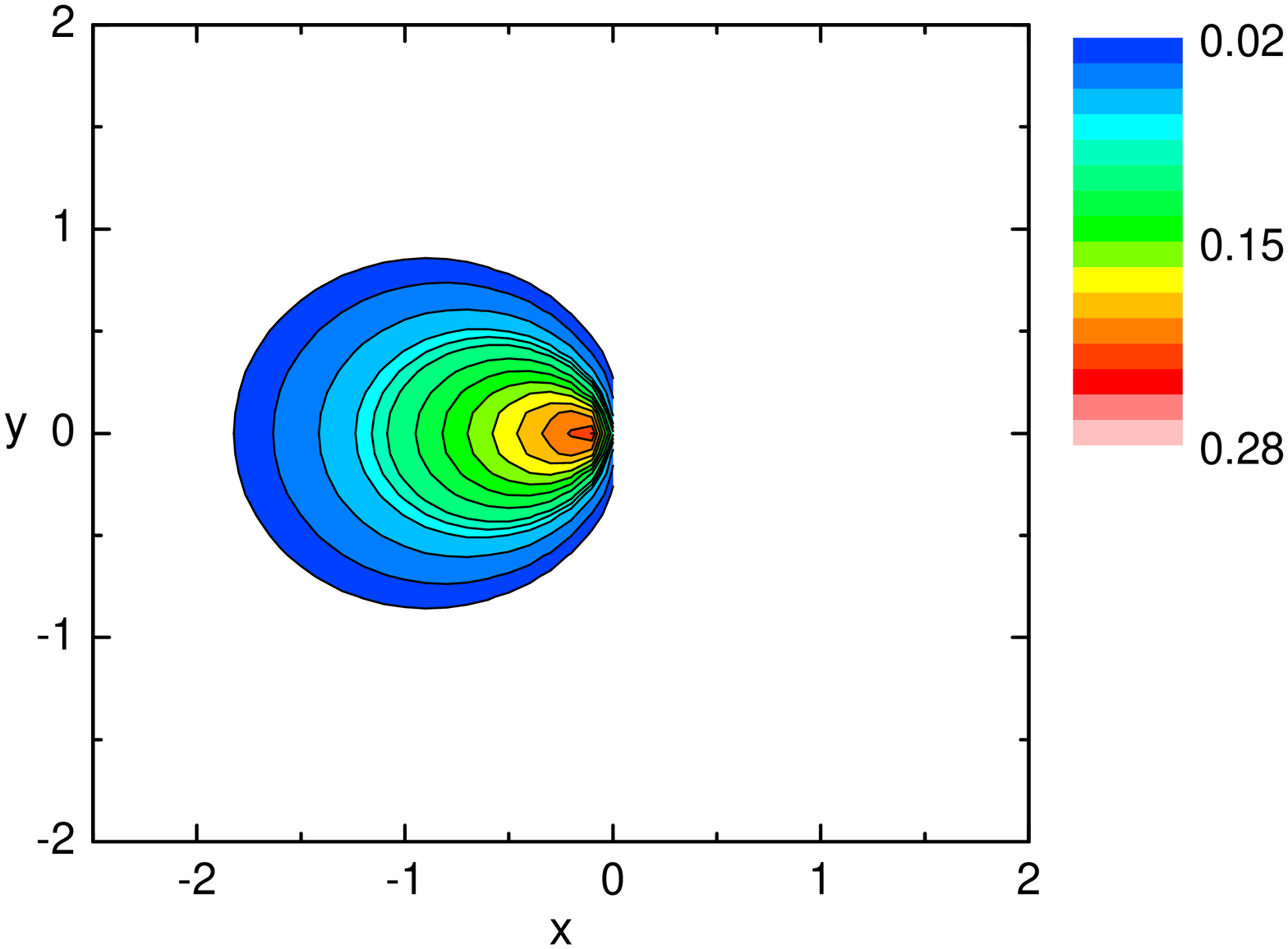}\\
\includegraphics[scale=0.29,angle=-90]{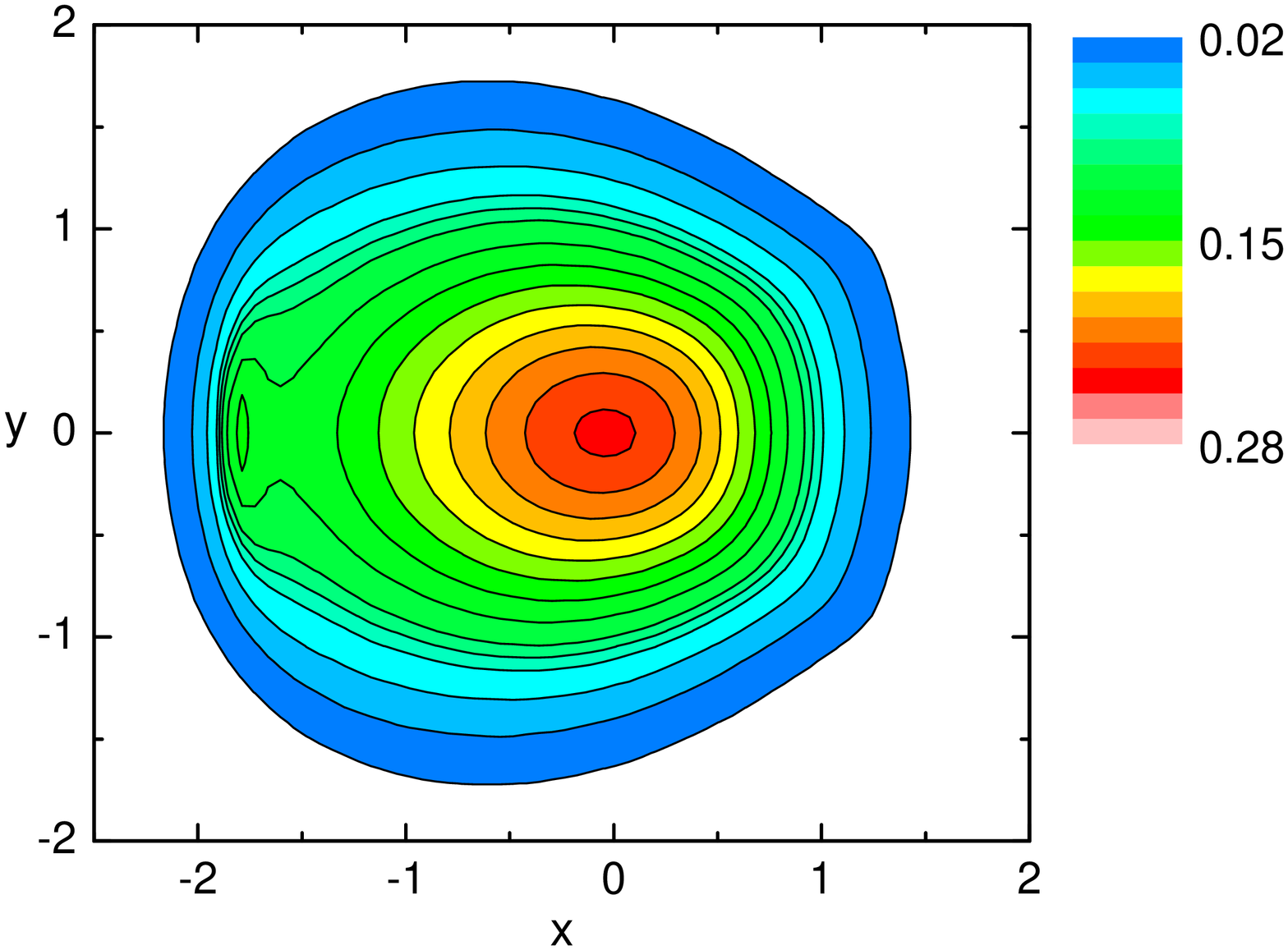}
\caption{(color online) Probability density at $T=2.4$ for the
quartic oscillator, with $\q=(0,0)$ and $\p=(2,0)$. The upper
panel shows the exact calculation, the middle one is
$|\psi_\q(x,y)|^2$ and the lower one is $|\psi_{\rm sc}(x,y)|^2$.
Using only real trajectories the result is very poor, but it
becomes excellent when complex ones are used: the overlap
(\ref{over}) in this case is around $95\%$.}
\end{figure}

\subsection{A bound system}

We now study a bound system, an isotropic quartic oscillator: \be
V(r)=Ar^2+Br^4,\ee where $A=0.5$ and $B=0.1$. The initial wave
packet has parameters $b_x=b_y=1$, $\q=(0,0)$ and $\p=(2,0)$,
which corresponds to a classical initial condition that is
periodic with period $\tau\approx 4.7$. In Fig. 2a we show the
probability density at $T=2.4$, approximately half the classical
period. It has a main peak at the origin and a small shoulder
around $x\approx -1.5$.

It is interesting that the approximation $\psi_\q(x,y,T)$ becomes
discontinuous in this case, as we see in Fig. 2b. This happens
because only a certain region of coordinate space can be reached
by real trajectories that start in the initial point $\q$ with an
initial momentum that is close to $\p$. Points outside of this
region can eventually be reached, but the initial momentum must be
so different from $\p$ that the actual contribution to $\psi_\q$
is negligible. In the border of this region there is a caustic,
where the wave function diverges, and in the numerical simulations
we must make this region a little smaller in order to avoid this.
All approximations based on real trajectories suffer from this
shortcoming, except for the IVR $\psi_{\q\p}$, which is always
Gaussian.

\begin{figure}[t]
\includegraphics[scale=0.28,angle=-90]{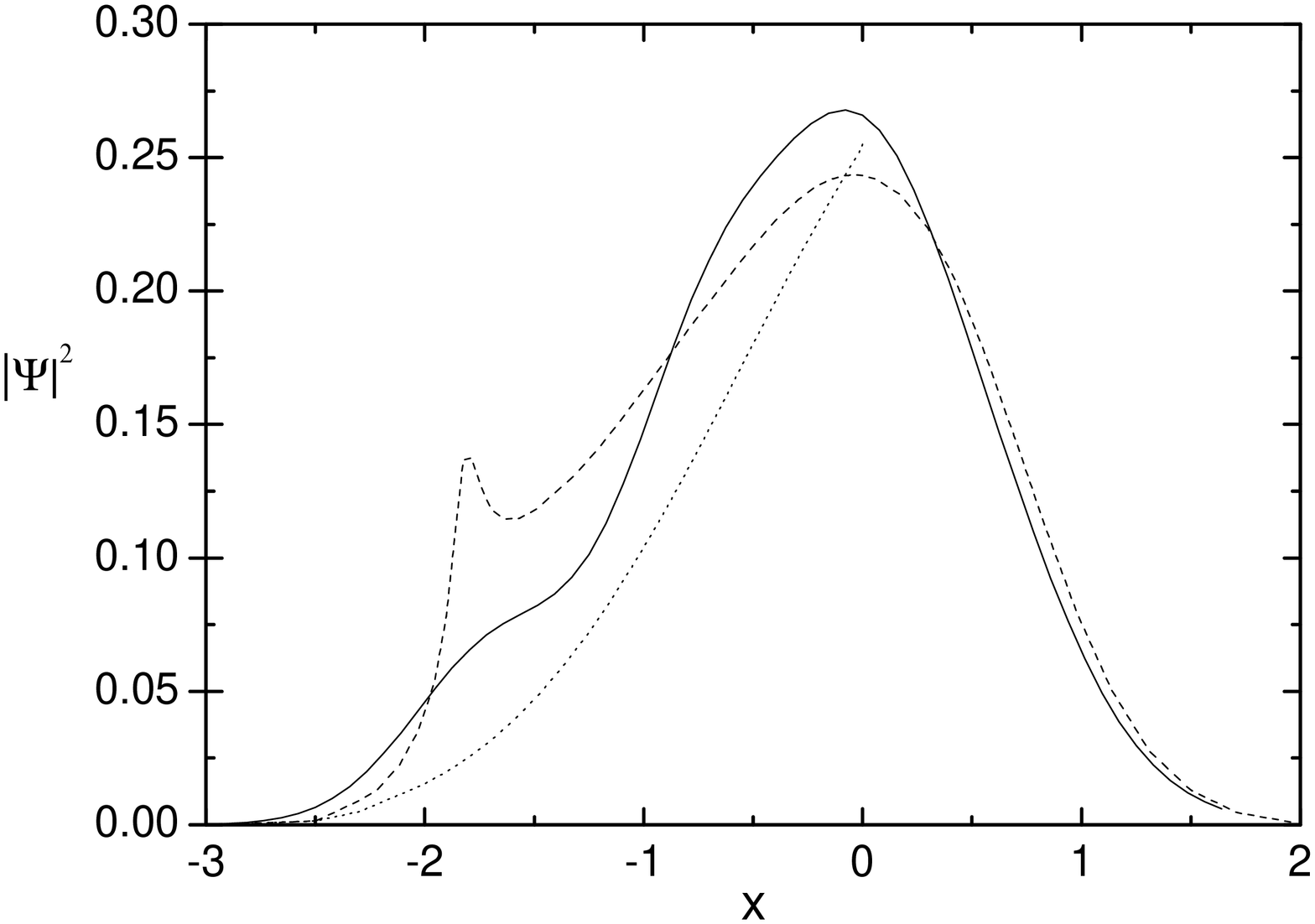}
\caption{Cut of the probability densities in Fig.2 along the line
$y=0$. The solid line is the exact result, the dashed line is
$|\psi_{\rm sc}(x,y)|^2$ and the dotted line is
$|\psi_{\q}(x,y)|^2$. Notice that the latter must be cut because
of the presence of a caustic.}
\end{figure}

The approximation $\psi_{\rm sc}$, on the other hand, is based on
complex trajectories and is well behaved in this case. It is
presented in Fig. 2c, where we can see that it reproduces very
well the main peak. In fact, its only defect occurs near the
shoulder. This is so because we have used only the main family,
and in that region a contribution from a secondary family should
be taken into account in order to give a good approximation (a
similar effect can be observed in a one dimensional quartic
oscillator \cite{aguiar}, where finding the secondary family is
much easier). It is interesting that in this case $\psi_{\rm sc}$
is far superior to the simpler real trajectory approximations, in
contrast with the previous example where no caustics appeared, and
the overlap (\ref{over}) in this case is around $95\%$.

A better picture of the behavior of these wave functions is given
in Fig. 3, where we show a cut along the line $y=0$ of the
previous plots. The exact probability density is the solid line,
while $|\psi_{\rm sc}(x,y)|^2$ is the dashed line and
$|\psi_{\q}(x,y)|^2$ the dotted line. The first two agree well
except around the region where the exact calculation has a small
shoulder. Inclusion of other families would certainly improve this
result. As already noted, the approximation based on the real
trajectory $\q\to\x$ fails completely for positive $x$ because of
the presence of a caustic line.

\subsection{Circular billiard}

As our third example, we consider the motion inside a circular
billiard with hard walls. If the particle is initially at the
center of the circle, the classical trajectories and also the
tangent matrix can be computed analytically, and we therefore
consider this case only. An exact calculation for $T=0.5$ is
presented in Fig. 4a, where we have used $\p=(4,0)$ and the radius
of the billiard is $R=3$ (once again we use $b_x=b_y=1$). As the
packet approaches the wall, it develops interference fringes in
the radial direction.

\begin{figure}[b]
\includegraphics[scale=0.29,angle=-90]{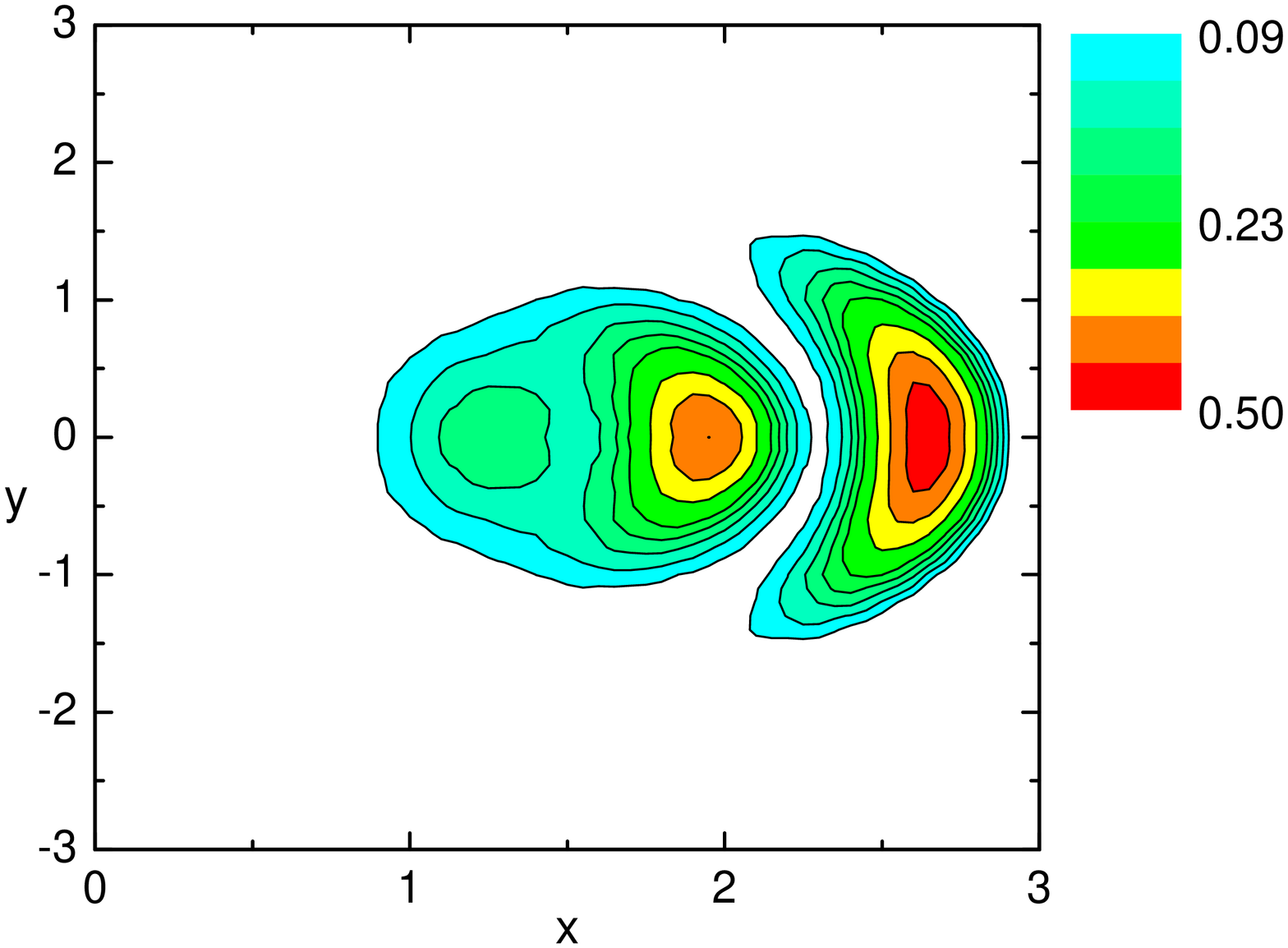}\\
\includegraphics[scale=0.29,angle=-90]{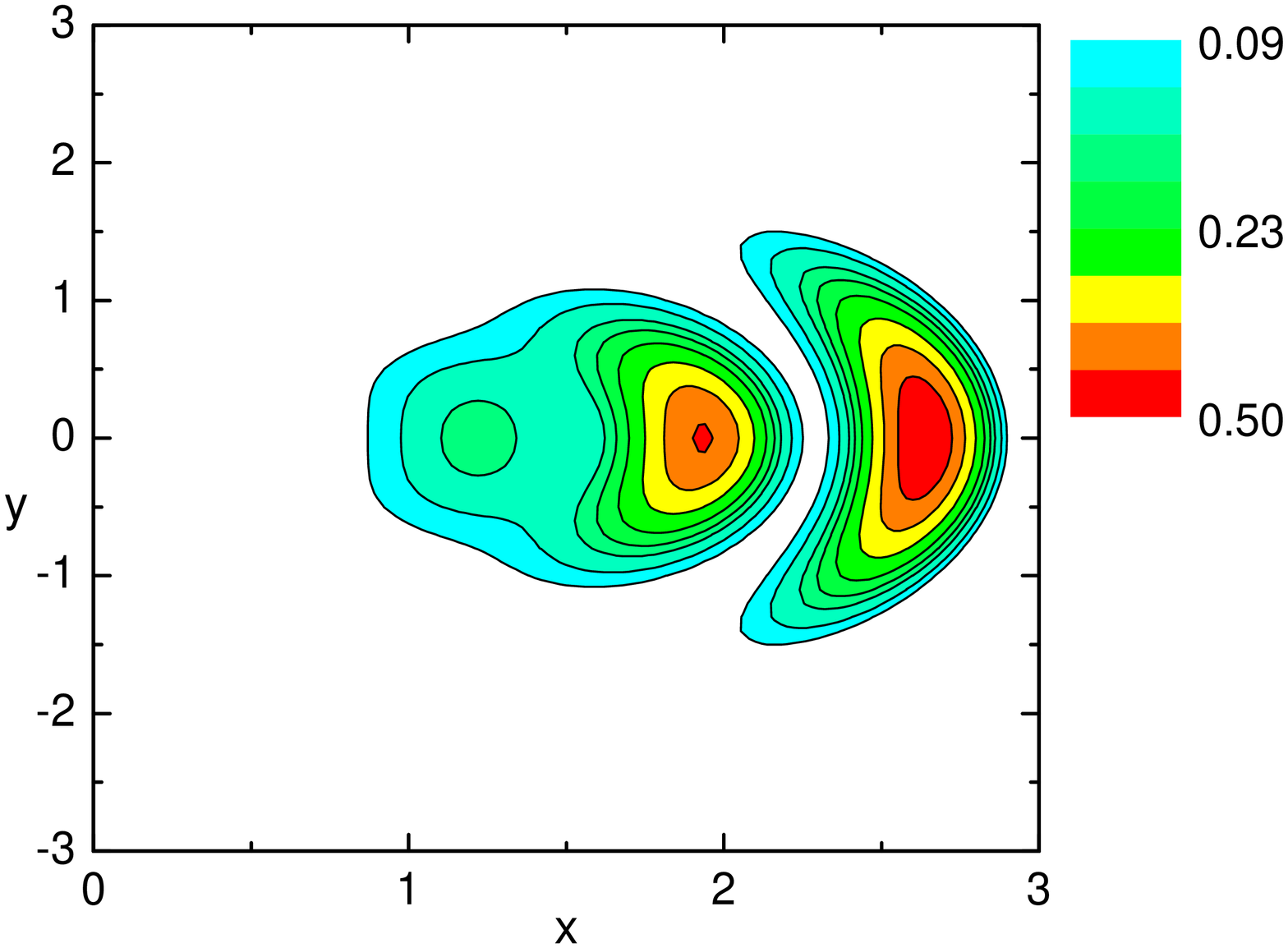}
\caption{(color online) Probability density at $T=0.5$ in the case
of a circular billiard, with $\q=(0,0)$ and $\p=(4,0)$. The upper
panel shows the exact calculation and the lower one is
$|\psi_{\q}(x,y)|^2$. Using only real trajectories we have a very
good result ($|\langle \psi|\psi_{\q}\rangle|^2\approx 97\%$),
including effects due to curvature and interference.}
\end{figure}

\begin{figure}[t]
\includegraphics[scale=0.28,angle=-90]{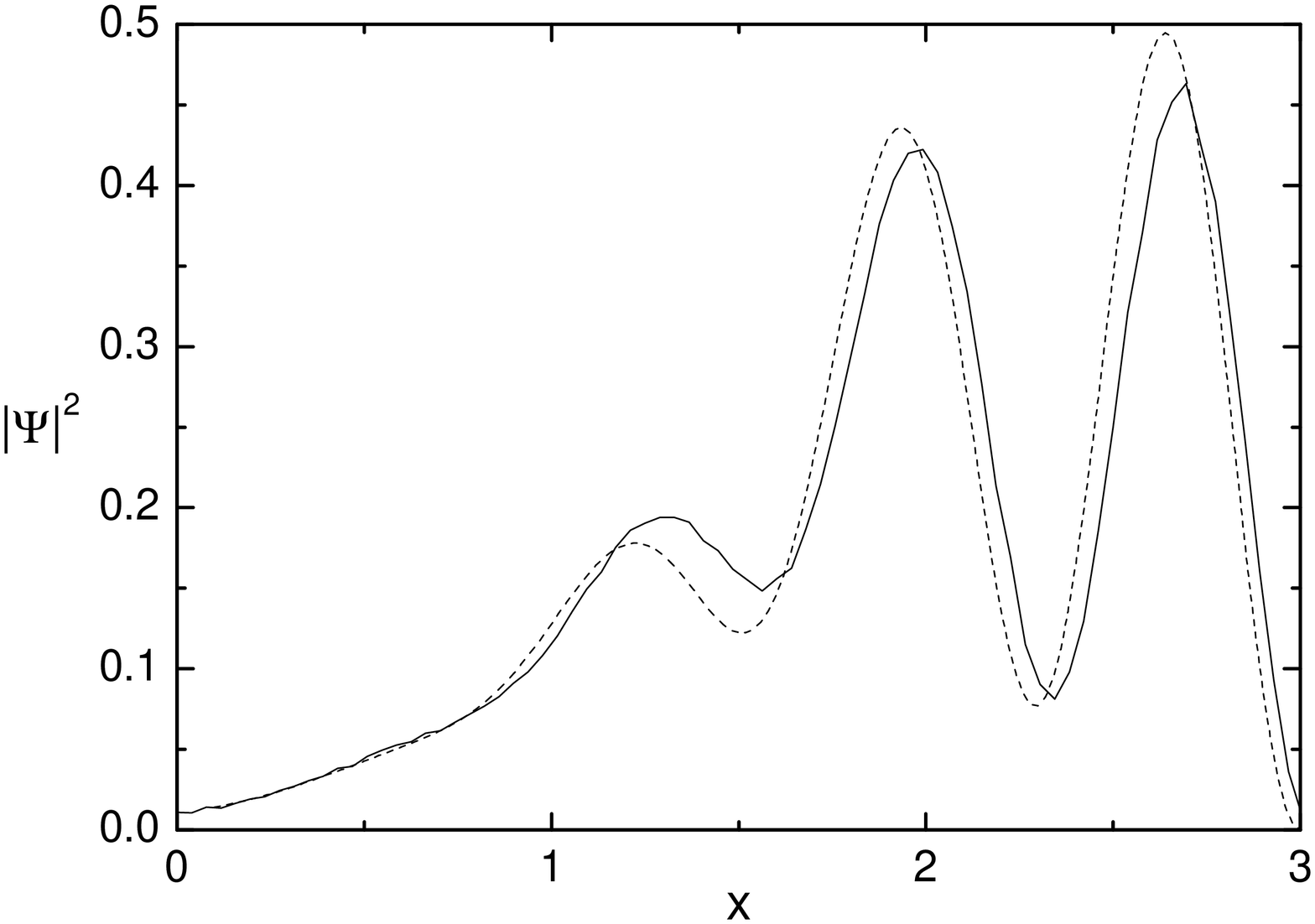}
\caption{Cut of the probability densities in Fig.4 along the line
$y=0$, displaying the exact (solid) and the semiclassical (dashed)
results. The latter is obtained from the interference of a direct
and a reflected trajectory.}
\end{figure}

We consider only the real approximation $\psi_\q$, but in this
case for all final points $\x$ we should take into account the
contribution of the many trajectories that reflect at the
boundaries of the billiard. The actual number of such trajectories
is infinite, but we consider only the two shortest ones,
respectively with zero and one reflection, which give the main
contributions. This gives origin to interference, as we can
appreciate from Fig. 4b. The agreement with the exact result is
excellent: the curvature is practically the same, as well as the
height and the position of the peaks. It is important to note that
there is a collision with a hard wall involved, and thus an extra
phase of $\pi/2$ must be introduced in the contribution of the
reflected trajectory. The overlap (\ref{over}) in this case is
around $97\%$. Since this approximation is already very good, we
do not present the complex calculation. We show again a cut along
the line $y=0$ of the probability densities, in Fig. 5. The small
discrepancy could be corrected if a twice reflected trajectory was
included.

\subsection{Tunnelling system}

Finally we consider a system in which the tunnel effect plays an
important role. We take a potential of the type \be
V(x,y)=V_0\exp\{-\frac{(r^2-r_0^2)^2}{\sigma^2}\},\ee with
$r^2=x^2+y^2$, which describes a circular ridge of radius $r_0$ in
the plane, centered around the origin. When an incident wave
packet with energy less than $V_0$ is scattered by this potential,
there is a probability that the particle will tunnel into the
ridge. In Fig. 6a we see the exact calculation at $T=2.5$, for a
potential with $V_0=10$, $r_0=5$, $\sigma=10$ and an initial wave
packet with $\q=(-10,0)$ and $\p=(4,0)$. The total probability of
being located inside the ridge is around $10\%$ in this case.

A semiclassical calculation for tunnelling though a square barrier
involving complex trajectories was presented in \cite{xavier},
where only the coherent state representation $\langle
z'|e^{-iHT/\hbar}|z\rangle$ was considered. In the present case
the classical motion must be solved numerically and the presence
of turning points leads to the appearance of caustics.
Nevertheless, provided the probability amplitude is not large in
the vicinity of the caustics, the real trajectory approximation
$|\psi_{\q}(x,y)|^2$ is able to give an accurate result, as we can
see in Fig. 6b (the overlap between the transmitted wave function
in the exact and semiclassical calculations is around $94\%$).
This is easy to understand if we remember that for each value of
the pair $(x,y)$ we need a different initial momentum $\p_i$ and,
even though a classical particle with the average momentum $\p$
would be reflected by the potential, there will be values of
$\p_i$ for which transmission is possible. The other real
trajectory approximation $|\psi_{\p}(x,y)|^2$, on the other hand,
works poorly in this case, because it involves variation only on
the initial position and this does not affect the energy of the
trajectories.

\begin{figure}[t]
\includegraphics[scale=0.29,angle=-90]{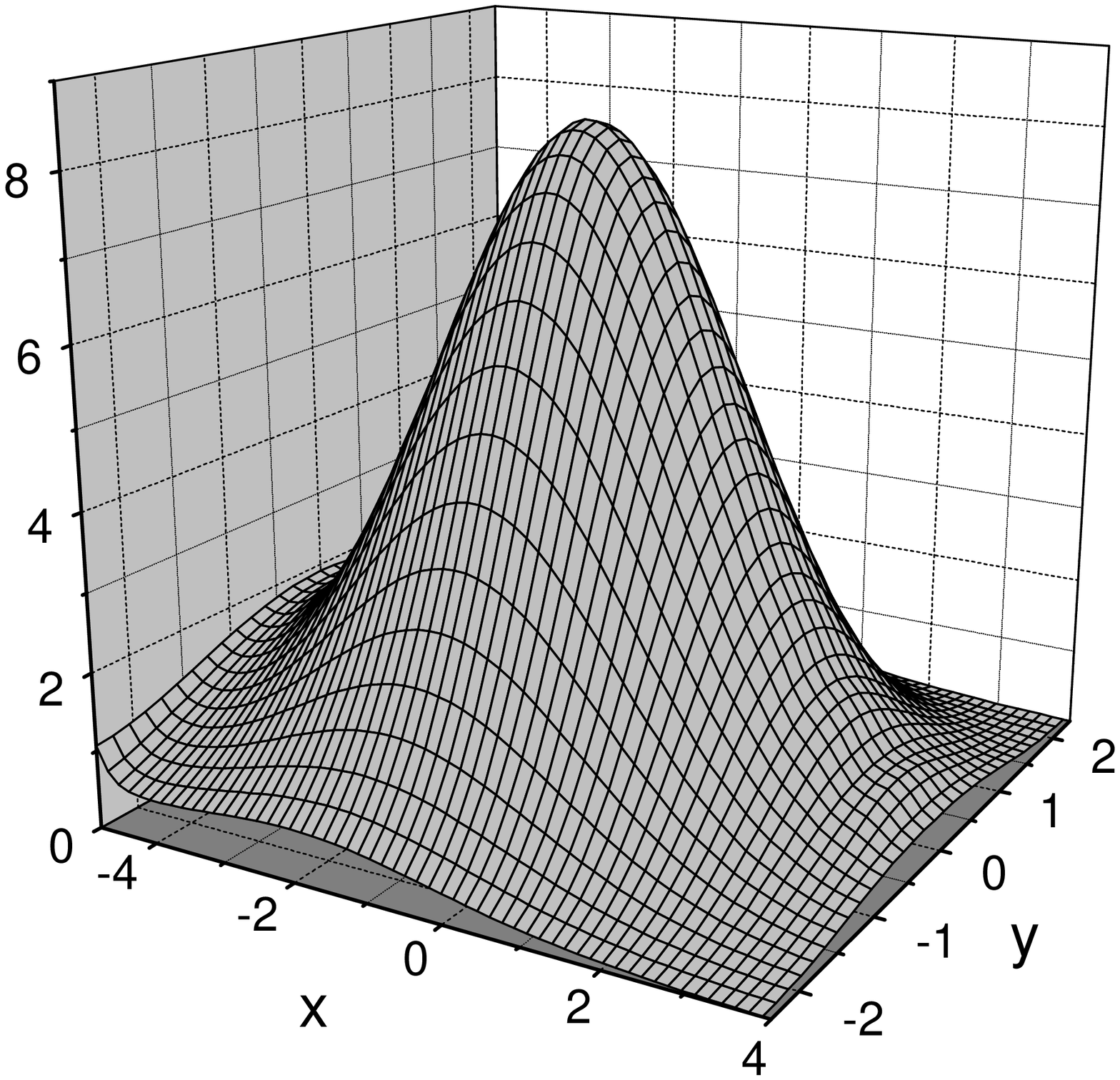}\\
\includegraphics[scale=0.29,angle=-90]{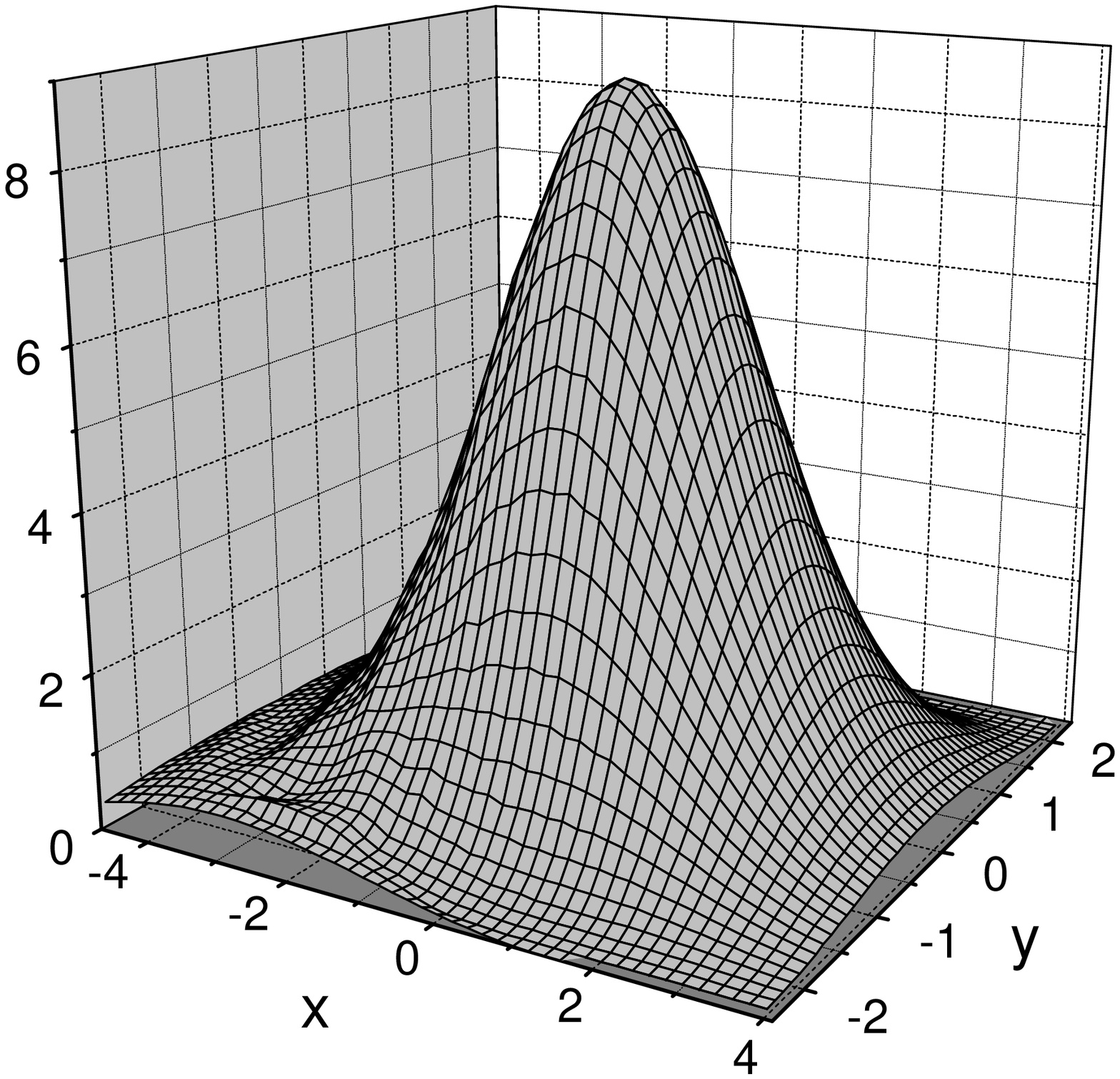}
\caption{Probability density at $T=2.5$ for the ridge potential,
with $\q=(-10,0)$ and $\p=(4,0)$. The upper panel shows the exact
calculation and the lower one is $|\psi_{\q}(x,y)|^2$ (times
$10^2$). Using only real trajectories it is possible to accurately
reproduce tunnelling effects ($|\langle
\psi|\psi_{\q}\rangle|^2\approx 94\%$).}
\end{figure}

The full complex semiclassical calculation would give even better
results than Fig. 6b, but this requires extending the potential to
the complex plane. This extension involves trigonometric functions
that make the numerical evolution very demanding. It is clear that
the simplicity of the trajectories involved in the calculation of
$|\psi_{\q}(x,y)|^2$ is of great practical advantage.

\section{Conclusion}

We have generalized the semiclassical approximation for the
propagation of wave packets based on complex trajectories derived
in \cite{huber,aguiar} to multidimensional systems. Several
further approximations based on real trajectories were also
derived from this basic formula, in particular Heller's Thawed
Gaussian Approximation (TGA). Apart from the TGA, all other
formulas are not Initial Value Representations and are able to
accurately reproduce non-Gaussian wave functions and also quantum
interference when more than one family of trajectories is present.

These theoretical results were tested in very distinct particular
cases, starting with scattering by an attractive potential, where
the classical trajectories must be computed numerically. For
positive energies this potential has no turning points and thus no
caustics. The complex and real approximations give
indistinguishable results that are very close to the exact
calculation. The second case was a bound nonlinear system, where a
large number of contributing classical trajectories exist. Using
only the main family we obtained a very good result with the
complex approximation. In this case the real trajectories
approximations are not practical because of the many caustics
involved. We also studied the motion inside a circular billiard,
taking into account two real trajectories for $\psi_\q(\x,T)$,
which displayed effects of curvature and interference. Finally, we
considered the tunnel effect and showed that again $\psi_\q(\x,T)$
is able to accurately reproduce the quantum result.

All cases studied in this paper are integrable and have circular
symmetry, which clearly introduces simplifications. We have also
considered relatively short propagation times. For long times the
number of trajectories in bound systems increases and caustics
proliferate, making a practical application of the formulas more
difficult. If it is possible to overcome this problem, the study
of chaotic systems would naturally be the next step.

\begin{acknowledgments}
Financial support from the Brazilian agencies FAPESP and CNPq is
gratefully acknowledged. We also thank Michel Baranger for
important discussions and an anonymous referee for useful remarks.
\end{acknowledgments}

\appendix*
\section{}

Consider a classical trajectory, satisfying Hamilton's equation
\be \frac{d}{dt}\begin{pmatrix} \x \\ \p \end{pmatrix}=J\nabla
H,\ee where $J$ is the usual symplectic matrix and $\nabla$ is the
$2d$-dimensional gradient. A variation around this trajectory satisfies
\be\label{a2}  \frac{d}{dt}\begin{pmatrix} \delta\x \\
\delta\p
\end{pmatrix}=J\begin{pmatrix}
  H_{\x\x} & H_{\x\p} \\
  H_{\p\x} & H_{\p\p}
\end{pmatrix}\begin{pmatrix} \delta\x \\
\delta\p
\end{pmatrix},\ee where the second derivatives of $H$ are computed at the reference trajectory.
Multiplying both sides on the left by a matrix containing the
inverse quantum uncertainties, $B$ and $C$, and inserting an
identity in the r.h.s.
we can rewrite (\ref{a2}) as \be\label{hqq}\frac{d}{dt}\begin{pmatrix}\delta\tilde{\x} \\
\delta\tilde{\p}
\end{pmatrix}=\begin{pmatrix}
  B^{-1}H_{\p\x}B & B^{-1}H_{\p\p}C \\
  -C^{-1}H_{\x\x}B & -C^{-1}H_{\x\p}C
\end{pmatrix}\begin{pmatrix} \delta\tilde{\x} \\
\delta\tilde{\p}
\end{pmatrix},\ee where $\tilde{\x}=B^{-1}\x$ and
$\tilde{\p}=C^{-1}\p$.

Now consider a trajectory that starts from $\x'$ with momentum
$\p'$ and arrives at $\x$ with momentum $\p$ (not related with the
initial coherent state label), and suppose we make small
displacements in its initial and final coordinates. This induces
variations in the initial and final momenta according to
\begin{equation}\label{disp}
\left(\begin{array}{c}
  \delta \p \\
  \delta \p'
\end{array}\right)=\begin{pmatrix}
   S_{\x\x} &  S_{\x\x'} \\
   -S_{\x'\x} &  -S_{\x'\x'} \
   \end{pmatrix}\left(\begin{array}{c}
     \delta\x\\
     \delta\x'
   \end{array}\right).
\end{equation}
On the other hand, the tangent matrix is defined to be the linear
application that relates the initial and final displacements,
\be\label{tangent}
\begin{pmatrix}
  \delta \tilde{\x} \\
  \delta \tilde{\p}
\end{pmatrix}=\begin{pmatrix}
  M_{\x\x} & M_{\x\p} \\
  M_{\p\x} & M_{\p\p}
\end{pmatrix}\begin{pmatrix}
  \delta \tilde{\x}' \\
  \delta \tilde{\p}'
\end{pmatrix},\ee where we have included explicitly the
quantum uncertainties for convenience. Inverting equation
(\ref{disp}) it is possible to show that \be
|-S_{\x'\x}^{-1}|=\frac{|B|}{|C|}|M_{\x\p}|=\frac{|B|^2}{\hbar^d}|M_{\x\p}|,\ee
which we have used in equation (\ref{pref}). It is also possible
to show that $S_{\x'\x'}=CM_{\x\p}^{-1}M_{\x\x}B^{-1}$ and
therefore \be
\Phi_{\x'\x'}=B^{-1}(iM_{\x\p}^{-1}M_{\x\x}-1)B^{-1},\ee where we
have used $C/\hbar=B^{-1}$. This leads to \be\label{det2}
|\Phi_{\x'\x'}|=\frac{i^d}{|B|^2}\frac{|M_{\x\x}+iM_{\x\p}|}{|M_{\x\p}|},
\ee as stated in equation (\ref{det}). The inverse of
$\Phi_{\x'\x'}$, used in (\ref{mist}), can also be expressed in
terms of the tangent matrix: \be
\Phi_{\x'\x'}^{-1}=-iB(M_{\x\x}+iM_{\x\p})^{-1}M_{\x\p}B.\ee

If we now take the time derivative of equation (\ref{tangent}),
and compare the result with (\ref{hqq}) we conclude that \be
\frac{dM}{dt}=\begin{pmatrix}
  B^{-1}H_{\p\x}B & B^{-1}H_{\p\p}C \\
  -C^{-1}H_{\x\x}B & -C^{-1}H_{\x\p}C
\end{pmatrix}M.\ee This is the dynamical equation for the tangent
matrix, which may be simplified for the large number of cases in
which $H_{\x\p}=H_{\p\x}=0$ and $H_{\p\p}$ is the inverse of the
mass. In practical applications these may be solved together with
the equations of motion, making it possible to follow the phase of
the prefactor in (\ref{final}).

\end{document}